\documentclass[sigconf, nonacm]{acmart}

\usepackage{amsmath}
\usepackage{tabularx, threeparttable, booktabs}
\usepackage{graphicx} 
\usepackage{listings}
\usepackage{cleveref}
\usepackage{tikz}

\keywords{Code Comprehension, Model Evaluation and Benchmarking, Machine Learning for Software Engineering}

\ccsdesc[500]{Computing methodologies~Machine learning}
\ccsdesc[500]{Software and its engineering}

\title{Beyond Accuracy: Characterizing Code Comprehension Capabilities in (Large) Language Models}

\newcommand{\affUzL}{%
  \affiliation{%
    \institution{ITS, University of Luebeck}
    \city{Luebeck}
    \country{Germany}
  }%
}
\author{Felix Mächtle}
\orcid{0009-0009-2431-0322}
\affUzL

\author{Jan-Niclas Serr}
\orcid{0009-0001-4624-8592}
\affUzL

\author{Nils Loose}
\orcid{0009-0003-6243-1623}
\affUzL

\author{Thomas Eisenbarth}
\orcid{0000-0003-1116-6973}
\affUzL

\renewcommand{\shortauthors}{Mächtle, et al.}

\date{}

\begin{document}

\begin{abstract}
Large Language Models (LLMs) are increasingly integrated into software engineering workflows, yet current benchmarks provide only coarse performance summaries that obscure the diverse capabilities and limitations of these models. 
This paper investigates whether LLMs’ code-comprehension performance aligns with traditional human-centric software metrics or instead reflects distinct, non-human regularities.
We introduce a diagnostic framework that reframes code understanding as a binary input–output consistency task, enabling the evaluation of classification and generative models. 
Using a large-scale dataset, we correlate model performance with traditional, human-centric complexity metrics, such as lexical size, control-flow complexity, and abstract syntax tree structure. 
Our analyses reveal minimal correlation between human-defined metrics and LLM success (AUROC~0.63), while shadow models achieve substantially higher predictive performance (AUROC~0.86), capturing complex, partially predictable patterns beyond traditional software measures. These findings suggest that LLM comprehension reflects model-specific regularities only partially accessible through either human-designed or learned features, emphasizing the need for benchmark methodologies that move beyond aggregate accuracy and toward instance-level diagnostics, while acknowledging fundamental limits in predicting correct outcomes.
\end{abstract}

\maketitle

\section{Introduction}
\label{sec:introduction}

Large Language Models (LLMs) have profoundly reshaped software development, facilitating tasks from automated code generation over debugging support to vulnerability classification. As LLMs become integral to developers' workflows, accurately assessing their strengths and weaknesses across various coding tasks has grown essential. Typically, developers depend on benchmark results to guide model selection, viewing aggregate success metrics as reliable indicators of effectiveness. However, these aggregated metrics often conceal critical nuances in model performance.

Consider two software developers: Alice, who predominantly tackles complex algorithmic challenges requiring mathematical reasoning, and Bob, who primarily addresses web development. Both are interested in leveraging LLMs for productivity enhancement. When consulting standard benchmarks, such as SWE-bench~\cite{jimenez2024swebench}, they find performance metrics that aggregate success across diverse task sets. However, such generalized metrics provide limited guidance, obscuring which tasks or domains each model excels in.

\begin{figure}[t]
$\vspace{2em}$\\
\centering
\begin{tikzpicture}

  \useasboundingbox (-2,-1) rectangle (2,1);

  \definecolor{colA}{HTML}{44AA99} 
  \definecolor{colB}{HTML}{DDCC77} 

  \coordinate (A) at (-0.55,0);
  \coordinate (B) at (0.55,0);
  \def\RX{0.85}
  \def\RY{0.71}

  \fill[colA, opacity=0.7] (A) circle (1.0);
  \fill[colB, opacity=0.7] (B) circle (1.0);

  \draw[black, thick] (A) circle (1.0);
  \draw[black, thick] (B) circle (1.0);

  \node[anchor=west, align=left] at (0.55+\RX,0+\RY) {OpenHands +\\  DevStral Small};
  
  \node[anchor=east, align=left] at (-0.55-\RX,0+\RY) {PatchPilot + \\ Co-PatcheR};

  \node[anchor=center] at (-1.0,0) {59};
  \node[anchor=center] at (0.0,0) {175};
  \node[anchor=center] at (1.0,0) {55};
  
\end{tikzpicture}
\Description{The Venn diagram shows that OpenHands + DevStral Small solves 234 tasks and PatchPilot + Co-PatcheR solves 230 tasks, with 175 tasks solved by both.}

\caption{Overlap of solved benchmark tasks on the verified SWE-bench dataset by the top two open-source models. While overall performance differs minimally (just 4 unique tasks difference), their solved task types substantially diverge, demonstrating  differences in model specialization.}
\label{fig:swebench_overlap}
\end{figure}
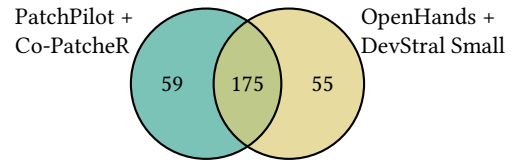

\Cref{fig:swebench_overlap}  illustrates this issue, showing the overlap in solved tasks between the two best-performing open-source models on the verified  SWE-bench dataset (commit ID 
\href{https://github.com/SWE-bench/experiments/tree/40c982da2f8608bdda4f272ff336c6da696d2639}{40c982d}). Although both models solve a similar total number of tasks, i.e., 234 tasks solved by model A (\textit{OpenHands + DevStral Small}) and 230 tasks by model B (\textit{PatchPilot + Co-PatcheR}), their strengths diverge. While 175 tasks overlap, model A uniquely solves 59 tasks and model B uniquely solves 55. This scenario implies that Alice might prefer model A if her tasks align closely with the unique strengths of that model, while Bob might favor model B due to its advantages. 

Motivated by this practical gap, our work investigates the foundations of LLM code comprehension: do models succeed or fail due to human‑interpretable code properties, or do they follow distinct, non‑human regularities? Specifically, we examine whether model performance correlates with traditional human-centric software metrics (e.g., line count, variable number, cyclomatic complexity) or whether it reflects different, latent regularities inaccessible to human interpretation. To this end, we design a diagnostic framework that reframes code understanding as a binary input–output (I/O) consistency task, enabling the evaluation of classification and generative models. Using a large-scale dataset derived from CodeNet~\cite{DBLP:conf/nips/Puri0JZDZD0CDTB21/CodeNet}, we compare the predictive power of two approaches, i.e., 
(i) a classifier trained on human-engineered static code features, and
(ii) a feature-free shadow model trained directly on raw code and I/O pairs.

Our analysis reveals that traditional software metrics correlate only weakly with model success (AUROC~0.63), suggesting that LLMs' comprehension capabilities diverge  from human reasoning about code. In contrast, shadow models achieve moderately higher predictive performance (AUROC~0.86), indicating that learned representations capture systematic patterns beyond handcrafted features, yet substantial prediction uncertainty remains. These findings highlight the need for new benchmarking methodologies that move beyond aggregate accuracy toward instance-level diagnostics, while acknowledging fundamental limits: LLM code comprehension reflects model-specific, partially predictable regularities that differ from human cognitive processes and resist complete characterization. All hyperparameters and implementation details are available in our GitHub Repository: \href{https://github.com/UzL-ITS/code-comprehension-capabilities-llms}{https://github.com/UzL-ITS/code-comprehension-capabilities-llms}.

\section{Preliminaries}
\label{sec:prelim}

\subsection{CodeNet}
CodeNet~\cite{DBLP:conf/nips/Puri0JZDZD0CDTB21/CodeNet} is an extensive dataset specifically created for research on code-related tasks. It consists of approximately 14 million code samples spanning over 50 programming languages, collected from popular online programming platforms.

\subsection{(Large) Language Models}
\label{sec:prelim:language-models}
Early code classification models, such as CNNs, RNNs, and graph-based networks, captured local or sequential code patterns but lacked long-range contextual understanding. Transformer-based encoders like UniXcoder~\cite{DBLP:conf/acl/GuoLDW0022/UniXcoder} addressed this through self-attention, enabling richer code representations across languages and tasks. More recently, decoder-only LLMs (e.g., GPT-4, CodeLlama, Mistral) have demonstrated strong generalization through large-scale pretraining and in-context reasoning.

\subsection{SAGE}
\label{sec:prelim:sage}
\emph{Shapley Additive Global importance} (SAGE)~\cite{DBLP:conf/nips/CovertLL20/SAGE} quantifies how much each feature contributes to a model’s predictive performance. Inspired by cooperative game theory, SAGE computes the expected reduction in loss when revealing a feature, averaged over all possible feature subsets. Thus, SAGE provides a global, dependence-aware measure of feature importance that accounts for correlations between inputs.

\section{Methodology}
\label{sec:method}

Our goal is to test whether a model's ability to understand code aligns with human-centric software metrics or instead reflect distinct,
non-human regularities. To this end, we cast code comprehension as a binary classification problem with ground-truth labels (\Cref{sec:method:task}), distill
a large set of static, execution-free metrics/features that summarize programs at the
source and bytecode levels (\Cref{sec:method:metrics}), and  present two predictors of LLM success: a classifier trained on human metrics (\Cref{sec:method:human-predictor}) and a feature-free shadow model trained directly on raw inputs (\Cref{sec:method:shadow}).

\subsection{Task formulation and label construction}
\label{sec:method:task}

To evaluate the code comprehension capabilities of language models, an appropriate task formulation is required. Prior work, such as Gu \emph{et al.}~\cite{gu2024cruxeval}, assesses this by providing programs with concrete inputs and verifying whether the model generates the correct output. However, since language models are also frequently applied to classification problems, we frame our evaluation as a classification.

Let $p$ denote a program (source code), $x$ an input, and $y$ a candidate
output.  We define the ground-truth label
\[
t(p,x,y) \;=\; \big[\, y = f_p(x) \,\big],
\]
where $f_p$ is the (deterministic) input–output mapping implemented by $p$.
Positive examples $(p,x,y)$ are obtained by executing $p$ on $x$ and recording
the resulting output. To generate matched negative examples without
changing the program or input distribution, we pair each $(p,x)$ with an
output $y'$ drawn from a different input for the same program
($y' = f_p(x')$ with $x'\neq x \land y'\neq y$), yielding $(p,x,y')$.  This
in-program shuffling preserves lexical, stylistic, and domain characteristics of
$p$ while producing semantically incorrect I/O pairs.

Given a model $M$, we query $M$ with $(p,x,y)$ and obtain a binary judgment
$\hat{t}_M(p,x,y)\in\{0,1\}$ (\emph{does $y$ match the output of $p$ on $x$?}).  We
say that $M$ succeeds on the sample if its judgment matches the
ground truth:
\[
s_M(p,x,y) \;=\; \big[\, \hat{t}_M(p,x,y) = t(p,x,y) \,\big] .
\]
All downstream analyses predict or explain $s_M$.

\subsection{Human-centric metrics}
\label{sec:method:metrics}

To probe alignment with traditional software measures, we compute a
feature vector $\phi(p,x,y)$ for each triple $(p,x,y)$ without executing the program.
Features fall into four families:

\begin{itemize}
  \item \textbf{Size / lexical.} Character and token counts, identifier and
  comment counts, simple name statistics, and lengths of $x$ and $y$
  (e.g., \emph{code\_chars}, \emph{num\_identifiers}, \emph{len\_input}).

  \item \textbf{Opcode statistics.} Statistics of Python bytecode
  opcode IDs and names extracted from the compiled form of $p$
  (e.g., number of opcodes, per-opcode frequencies, entropy).

  \item \textbf{AST / graph structure.} Metrics computed on the abstract
  syntax tree of $p$ and its derived graphs (e.g., node/edge counts,
  density, diameter, average shortest path length).

  \item \textbf{Control-flow / complexity.} Cyclomatic complexity, loop and
  branch counts, and basic-block statistics.
\end{itemize}

All features are obtained via static analysis of source or compiled bytecode. The full list of about 300  features can be seen in our Github Repository. 

\subsection{Predicting LLM success from human metrics}
\label{sec:method:human-predictor}

We train a supervised classifier $g_{\theta}$ on ($\phi(p,x,y), s_M$), i.e., the human metrics for a sample, together with information on whether the model was successful for that sample. Thus, we assess
how much human-centric metrics predict the success of a model. 

To obtain global, dependence-aware feature importances, i.e., which metrics lead to model success/failure, we compute feature importance based on SAGE~\cite{DBLP:conf/nips/CovertLL20/SAGE}. SAGE values are
used to (i) rank features and (ii) construct compact models by retaining the
smallest subset whose positive SAGE contributions account for $95\%$ of the
total positive mass. Comparing full vs.\ pruned models
reveals whether a small set of human metrics suffices to predict $s_M$.

\subsection{Shadow model (feature-free predictor)}
\label{sec:method:shadow}

Human-designed features may miss high-dimensional cues that influence LLM
behavior. We therefore train a shadow model $\tilde{g}_{\psi}$ that
predicts $s_M$ directly from raw inputs. Concretely, we fine-tune an encoder
for code (UniXcoder~\cite{DBLP:conf/acl/GuoLDW0022/UniXcoder}) on serialized sequences
\[
\big\langle ~p~\texttt{[SEP]}~x~\texttt{[SEP]~y~} \big\rangle,
\]
omitting handcrafted features. This model estimates whether a target
model will answer correctly for a given $(p,x,y)$ based solely on the code and
input/output-context, serving as feature-free predictor.

\subsection{Dataset construction}
\label{sec:method:dataset}

We focus on the Python portion of CodeNet~\cite{DBLP:conf/nips/Puri0JZDZD0CDTB21/CodeNet}. For each program
$p$, we synthesize inputs with a lightweight, type-aware fuzzer and execute
$p$ to collect outputs, producing $799{,}365$ triples $(p,x,y)$. To prevent redundant or contradictory training signals, we
collapse all program–input pairs that exhibit identical observable behavior for
a given sample. Finally, the dataset is partitioned by problem statement: every tenth
problem is assigned to the evaluation set, while the remaining problems form
the training pool.

To respect the context limits of models and ensure stable preprocessing, we apply length
filters of at most $5{,}000$ characters for code and $500$ characters for each
of input and output. Before filtering, $152{,}803$ pairs are assigned to the
training split and $12{,}584$ to the test split. After filtering, the final
corpora contain $148{,}243$ training and $12{,}563$ evaluation pairs.  All
static features in \Cref{sec:method:metrics} are computed from these filtered
sources. 

\section{Experiments}
\label{sec:experiments}

\subsection{Model Performance}

Following prior benchmarks~\cite{gu2024cruxeval, jimenez2024swebench}, we begin by evaluating the performance of all models on our task. For models requiring training, we reuse the set of models selected in OCEAN~\cite{DBLP:conf/acns/MachtleSLSE25}, along with their publicly available implementations and training routines.

\textbf{Experiment setup.}
We evaluate general-purpose LLMs and classic text baselines on the binary classification task described in \Cref{sec:method:task}. Models that require training are fit on the training split of our dataset, frozen API models are prompted to decide whether the output matches the program’s behavior on the given input. The prompt follows the style of CruXEval~\cite{gu2024cruxeval}, adapted to our classification format. We report performance on the held-out test split. Results are shown in \Cref{tab:eval:performance-per-model-in-accuracy}.

\textbf{Discussion.} We observe that model scale and pretraining quality play decisive roles: larger decoder-based models consistently outperform shallow text baselines. The largest model with reasoning capabilities, GPT-OSS 120B, represents the strongest system in our comparison.  Interestingly, the much smaller yet code-specialized UniXcoder ranks second, demonstrating that compact models trained on domain-specific data can achieve competitive results. Moreover, the dominance of Transformer-based architectures across all high-performing systems highlights their effectiveness in capturing the structural and semantic dependencies inherent in code. The remainder of the paper focuses on the subset of working models, i.e., those achieving an F1 score above 0.65.

%
%
\begin{table}[t]
    \centering
    \caption{Aggregated performance of different models on our benchmark dataset. “I.” denotes instruction-tuned variants. }
    \begin{tabularx}{\linewidth}{X|cccc} \toprule
    \textbf{Model} & \textbf{Accuracy} & \textbf{Precision} & \textbf{Recall} & \textbf{F1} \\ \midrule
CodeLlama 13B I. & 0.506 & 0.033 & 0.600  & 0.062 \\ 
Llama 3.3 70B I. & 0.738 & 0.514 & 0.931  & 0.662 \\ 
Mistral Small 24B I. & 0.744 & 0.556 & 0.892  & 0.685 \\ 
GPT-OSS 120B & \textbf{0.960} & \textbf{0.926} & \textbf{0.995}  & \textbf{0.959} \\ 
Phi-4 & 0.733 & 0.554 & 0.862  & 0.674 \\ 

\midrule
CNN & 0.500 & 0.500 & 0.728 & 0.593 \\
FNN + N-grams & 0.539 & 0.551 & 0.417 & 0.475 \\
FNN + TF-IDF & 0.511 & 0.533 & 0.179 & 0.268 \\
LSTM & 0.500 & 0.500 & 0.000 & 0.000 \\
UniXcoder & \textbf{0.815} & \textbf{0.790} & \textbf{0.860} & \textbf{0.823} \\
\bottomrule
    \end{tabularx}
    \label{tab:eval:performance-per-model-in-accuracy}
\end{table}


\subsection{Human Metrics}
\label{sec:human-metrics}

Having established per-sample success labels for each model, we investigate whether traditional human-designed code properties correlate with LLM performance. We train a supervised classifier on the static features described in \Cref{sec:method:metrics} to predict model success, then apply SAGE~\cite{DBLP:conf/nips/CovertLL20/SAGE} to identify which metrics matter most.

\textbf{Experiment Setup.}
We train an XGBoost classifier on an 80/20 stratified train/test split, with features $\phi(p,x,y)$ as input and the binary success label $s_M$ as target. To obtain global feature importance rankings, we compute SAGE values on the held-out split. Following the 95\% coverage criterion from \Cref{sec:method:human-predictor}, we retain the smallest feature subset whose positive SAGE contributions account for 95\% of total positive mass. We retrain XGBoost on this reduced set and report test-set AUROC for both full and pruned models.

\textbf{Results.}
Across all models, SAGE pruning retained only $16.8$ features on average (5.8\%). Despite this drastic reduction, pruned models achieved mean AUROC of 0.634 versus 0.554 for full-feature baselines, an improvement of $+0.080$. The most frequently selected features correlated with the code length, yet no single feature dominated. Top-ranked features typically accounted for 11--60\% of positive SAGE mass, with the remainder distributed across opcode frequencies, AST node counts, and graph structure metrics.

\textbf{Discussion.} 
While intuitive signals such as code length and input size appear among top features, the bulk of predictive power is fragmented across dozens of low-level indicators (opcode distributions, AST motifs) without a unifying pattern. Furthermore, feature rankings vary substantially across models: UniXcoder prioritizes lexical cues (e.g. \texttt{token\_count}), whereas GPT-OSS-120B emphasizes opcodes and graph density. This heterogeneity, coupled with only modest AUROC scores, indicates that classical software metrics correlate only weakly with LLM comprehension. If they do, the correlation is largely driven by code length, i.e., a modest negative correlation between program length and model success. The data supports that LLM capability on this task transcends traditional human-centric complexity measures.

\subsection{Non-Human Signal via a Shadow Model}
\label{sec:shadow-model}

Having shown that human-designed metrics weakly predict performance, we investigate whether \emph{non-human} signals can do better. We fine-tune UniXcoder~\cite{DBLP:conf/acl/GuoLDW0022/UniXcoder} as a feature-free shadow model to predict a target model's correctness on each $(p,x,y)$ triple from raw inputs, bypassing handcrafted features and probing for predictive cues beyond human intuition.

\textbf{Experiment Setup.}
For each target model, we train a dedicated shadow model on an 80/20 stratified train/test split of the corresponding success labels. The input is a concatenated sequence $\langle p, x, y \rangle$ tokenized. Following OCEAN~\cite{DBLP:conf/acns/MachtleSLSE25}, we train for a single epoch and report performance on the test-set.

\textbf{Results.}
Shadow models achieve mean AUROC of 0.859.
Individual performance varies by target model: GPT-OSS~120B (AUROC~0.834), Mistral~Small~24B (AUROC~0.872), Llama 3.3~70B (AUROC~0.849), UniXcoder (AUROC~0.878), and Phi-4 (AUROC~0.864).

\textbf{Discussion.}
The shadow model's substantial improvement over human metrics (mean AUROC~0.859 vs.\ 0.634) confirms that non-human signals exist and are learnable. Unlike the human-metric classifier, which struggles to integrate weak indicators, the shadow model directly encodes high dimensional interactions between code structure, I/O context, and target-model behavior. However, the moderate AUROC ($\sim$0.86) reveals limits: even feature-free learned representations leave uncertainty. Those limits suggest that LLM failure patterns contain both systematic components (captured by the shadow model) and context-dependent factors beyond what raw code and I/O reveal. The consistent AUROC across diverse target models, 
indicates these signals generalize within the shadow modeling framework, yet they remain model-specific. Moreover, learned representations remain opaque: while shadow models outperform human metrics, they offer no interpretable explanation of \emph{why} certain samples fail. Taken together, these results demonstrate that LLM code comprehension patterns transcend classical complexity measures but reflect model-specific, partially predictable regularities rather than deterministic or universal principles.

\section{Related Work}
\label{sec:related}

The number of benchmarks for code understanding is increasing: SWE-bench tasks models with resolving real GitHub issues across multi-file contexts~\cite{jimenez2024swebench}. CRUXEval assesses   execution reasoning via input-output prediction on short Python programs~\cite{gu2024cruxeval}. 
In contrast, we do not assess the quality of generated code. Instead, we evaluate binary model classifications over given programs, thereby shifting from coarse aggregate scores to capability-level diagnostics.

A second line of work leverages software engineering metrics to assess LLM-generated code. Sepidband~\emph{et~al.} study whether human-centric  metrics can proxy the correctness of generated code~\cite{DBLP:conf/compsac/SepidbandTWH25}. We, however, do not evaluate the quality of generated code. Instead, we compute human-centric metrics on the input program and use them to assess whether the model’s output classification is correct.

Complementary to our approach, Kim~\emph{et al}~\cite{DBLP:conf/icse-deeptest/KimKAY25} propose Lachesis,  
a framework that predicts whether large language model reasoning will yield a correct outcome by analyzing the structural properties of sampled reasoning paths. Using graph- and sequence-based representations, Lachesis anticipates inference success. 
While our shadow models likewise predict model outcomes, Lachesis targets reasoning-level reliability, whereas we focus on input-level code comprehension.

Closest to our length-based finding, Mächtle \emph{et al.} show that increasing input length degrades LLM performance on security-related tasks~\cite{maechtle2025tracegadgs}, and Rando \emph{et al.} report similar degradation on their novel code-comprehension benchmark~\cite{DBLP:journals/corr/abs-2505-07897/long-code-benchmark}. However, both analyses focus solely on length and omit other static and structural 
complexity indicators. Our results both corroborate this length effect and systematically broaden the metric space, while further introducing a feature-free shadow predictor that captures model-specific regularities beyond human-designed metrics.

\section{Conclusion and Outlook}

We introduce a diagnostic framework for assessing code comprehension in large language models that (i) reframes evaluation as a binary I/O-consistency judgment, (ii) relates performance to human-centric software metrics, and (iii) trains feature-free shadow predictors directly from raw code and I/O context. We found that classical measures, such as size, AST/CFG structure, or bytecode statistics, correlate only weakly with LLM success (AUROC~0.63). Shadow models trained on raw inputs achieve moderately higher predictive performance (AUROC~0.86), indicating that learned representations capture systematic patterns beyond handcrafted features.

These findings motivate three implications: 
First, aggregate benchmark scores obscure specialization: model selection should be task- and instance-aware rather than guided solely by headline accuracy. Second, the weak alignment between human metrics and LLM behavior underscores the opacity of current systems and motivates interpretability methods that move beyond handcrafted features. Third, the moderate shadow model performance (AUROC~0.86) highlights inherent unpredictability in LLM code comprehension, cautioning against over-reliance on predictive models 
in safety-critical applications. 

Future work must develop hybrid approaches that combine human-interpretable diagnostics with learned predictors while acknowledging irreducible uncertainty in LLM behavior.

\begin{acks}
This work has been
supported by funding from the Agentur für Innovation in der
Cybersicherheit GmbH (Cyberagentur, project SOVEREIGN)
\end{acks}

\bibliographystyle{plain}
\bibliography{References}

\end{document}